# Novel Cascaded Gaussian Mixture Model-Deep Neural Network Classifier for Speaker Identification in Emotional Talking Environments

*[1]Ismail Shahin, [2]Ali Bou Nassif, [3]Shibani Hamsa

Department of Electrical and Computer Engineering

University of Sharjah

P. O. Box: 27272

Sharjah, United Arab Emirates

[1]ismail@sharjah.ac.ae, [2]anassif@sharjah.ac.ae, [3]shibani.h@gmail.com

* Corresponding author

## Abstract

This research is an effort to present an effective approach to enhance text-independent speaker identification performance in emotional talking environments based on novel classifier called cascaded Gaussian Mixture Model-Deep Neural Network (GMM-DNN). Our current work focuses on proposing, implementing and evaluating a new approach for speaker identification in emotional talking environments based on cascaded Gaussian Mixture Model-Deep Neural Network as a classifier. The results point out that the cascaded GMM-DNN classifier improves speaker identification performance at various emotions using two distinct speech databases: Emirati speech database (Arabic United Arab Emirates dataset) and "Speech Under Simulated and Actual Stress (SUSAS)" English dataset. The proposed classifier outperforms classical classifiers such as Multilayer Perceptron (MLP) and Support Vector Machine (SVM) in each dataset. Speaker identification performance that has been attained based on the cascaded GMM-DNN is similar to that acquired from subjective assessment by human listeners.

## 1. Introduction

Speaker recognition and its sub divisional entities: "speaker identification and speaker verification" need to be redefined on the basis of talking environments as the neutral talking environment and the emotional talking environments. Speaker recognition performance faces drastic challenges, especially when the speaker identity going through the human-computer interface in emotional talking environments. Speaker recognition applications in security systems are widening their base into banking sector, customer care sector, criminal investigation and can be used as security control measure to remotely access a server or for access to confidential library files on a server. The process of "automatic speaker identification and verification" in stressful and emotional talking environments is a challenging area of research [1].

"Speaker identification" is comprised of two schemes in terms of sets: "closed set" and "open set" speaker identification. When the unknown speaker is presumed to be one among the database of known speakers, it becomes the scheme of a "closed set", while in the scheme of an "open set", the unfamiliar speaker might not necessarily be from the database of familiar speakers. Operational procedure divides speaker identification into "text-dependent", where the same text is uttered by the speaker in the training and testing phases and "text-independent", where different texts are uttered by the speaker during the training and testing phases [2].

A perfect communication from a speaker depends not only on linguistic statements but also on the emotional aspects of the speaker. Identifying the emotional aspects of the speaker by the machine is still a challenge of the human-machine interface. Speech is always a perfect mix of linguistic notes linked with emotion along with its paralinguistic features. Emotion recognition from speech is quite hard issue to solve, occasionally even a human fails to categorize spontaneous emotions based on the given speech signal. Hence, "speaker identification" in emotional environments is a challenging task [3,4,5]. This research studies a novel architecture based on both "Gaussian Mixture Model" and "Deep Neural Network" for "speaker identification" in different aspects of emotional environments.

This paper is organized as follows: Literature review is given in Section 2. The details of the Emirati Speech Database (ESD) are given in Section 3. Section 4 contains the details of feature extraction. Section 5 is organized with the details of classification. Section 6 explains the proposed algorithm using GMM-DNN classifier. Section 7 demonstrates the results obtained and further experiments conducted. Finally, concluding remarks are given in Section 8.

## 2. Literature Review

The importance of emotion recognition together with speaker identification has a great importance in the field of machine learning to bridge the gap between human computer interaction and intelligent human computer interaction. Some papers have spotlighted on studying speaker identification in emotional environments [6,7,8]. Li *et.al* [6] proposed a speech emotion-state transformation to improve speaker identification in enthusiastic talking circumstances. Wu *et.al* [7] described speaker identification in emotional environments by utilizing certain standards for modifying the feature element extraction techniques. Bao *et.al* [8] utilized two strategies to recognize the speaker in enthusiastic talking condition. The primary strategy is the Emotion Attribute Projection (EAP) and the second technique is a linear combination over "GMM-UBM based framework" and the "support vector machine".

Koolagudi [9] introduced a method of "speaker identification" using emotional environments by utilizing the suitable transformations of "Mel Frequency Cepstral Coefficients (MFCCs)" descriptors. Jawarkar [10] illustrates "text-independent speaker identification" involving five basic emotional environments, such as "anger, fear, happy, disgust and sad". He used a feature comparison method and a hybrid classifier model for the effective classification. "Text-independent speaker identification" was performed and tested based on GMM and SVM as classifiers with the Berlin database which is made up of five male and five female speakers talking in neutral, sad, happy, disgust, fear and bored emotions. Jawarkar tested and evaluated the rate of different feature groups such as: "MFCCs, Line Spectral Frequencies (LSFs), Teager energy based Mel Frequency Cepstral Coefficient (TMFCCs), Temporal Energy of Sub-band Cepstral Coefficient (TESBCCs) and their combinations MFCCs-LSFs and TESBCCs-LSFs" in text-independent emotional environments. Mansour *et.al* [11] considered MFCC-Shifted Delta Coefficient (SDC) features to improve the performance of "speaker recognition system in emotional talking environments".

Shahin suggested, executed and assessed a two-phase architecture for identifying speakers utilizing their passionate signals by consolidating a single-phase acknowledgment framework for the recognition of both speaker and emotions based on both HMM and SPHMM as classifiers [12]. Shahin enhanced the results by utilizing the two-phase approach. In another work, Shahin utilized gender signals along with the emotional prompts and speaker signals (three-stage framework) for better outcomes. The outcomes demonstrate that the three-stage framework is more successful than the two-stage framework [13].

Most of the work in the "speech and speaker recognition" areas is focused on speech uttered in the English language. Very few studies focused on these two areas using the Arabic language. Shahin and Ba-Hutair introduced and analyzed speaker identification of Emirati database using different classification techniques such as VQ, GMM and HMM [14]. Shahin used HMM, CSPHMM2 and SPHMM classifiers to increase the "speaker identification performance in such talking environments" [12,15,16].

Trigeorgis *et.al* proposed an approach to the problem of "context-aware" emotional related feature extraction based on merging "Convolutional Neural Network (CNN)" with "Long Short Term Memory (LSTM)" network to automatically learn the optimum illustration of unprocessed speech signals [17]. They demonstrated that their topology remarkably leads the classical methods using signal processing frameworks. Schmidt and Kim employed a "regression-based deep belief networks" to learn features directly from magnitude spectra for the application of music emotion recognition [18]. Matejka *et.al* studied utilizing "Deep Neural Network (DNN) Bottleneck (BN)" features jointly with the conventional "MFCC features" for "speaker recognition" [19]. They showed 60% relative gain competed to the classical MFCC baseline for EER, resulting in 0.94% EER. Lee *et.al* applied "convolutional deep belief network" to audio signals and experimentally assess their feature learning on distinct audio classification tasks [20]. They showed that the learned features belong to phones/phonemes in the case of speech signals. Richardson *et.al* proposed a method for "speaker recognition and language recognition" using a single DNN [21]. Their proposed approach was tested on the "2013 Domain Adaptation Challenge speaker recognition (DAC13) and the NIST 2011 Language Recognition Evaluation (LRE11)". Ali *et.al* proposed a framework of combining the learned features and the MFCC features for "speaker recognition" which can be implemented to audio scripts of various lengths [22]. Specifically, they studied utilizing features from distinct levels of "Deep Belief Network (DBN)" to quantize the audio data into vectors of audio word counts. They demonstrated in their research that the audio word count vectors produced from combination of DBN features at distinct layers yield higher rate than the MFCC features [22].

Support Vector Machine (SVM) has been widely used as a classifier in speaker recognition area. Nijhawan and Soni proposed text-dependent speaker recognition using SVM and obtained average speaker recognition rate of 95.0% [23]. Katz *et.al* investigated two discriminative classification frameworks for frame-based speaker identification and verification based on SVM and Sparse Kernel Logistic Regression (SKLR) [24]. In their work, they showed that both frameworks are superior to the GMM baseline for speaker identification and verification.

Multilayer Perceptron has been extensively used in the field of speaker recognition as a classifier. Sharma *et.al* proposed text-independent speaker identification using back propagation MLP network classifier for a closed set of speakers [25]. Srinivas *et.al* introduced a neural network based classification for speaker identification and wavelet transform for feature extraction [26]. Their introduced classifier was trained to designate a speaker name as a tag to the test speech signals.

In the present research, we aim at improving text-independent speaker identification performance in emotional talking environments based on a novel classifier called cascaded Gaussian Mixture Model-Deep Neural Network (GMM-DNN). This proposed classifier is composed of a hybrid Gaussian Mixture Model followed by a Deep Neural Network. Our novel classifier has been evaluated on two distinct and independent speech datasets: Arabic Emirati-accented dataset and English SUSAS database. Furthermore, four diverse tests have been conducted to assess our proposed classifier.

## 3. Emirati Speech Database

In this work, "25 male and 25 female local Emirati" speakers with ages spanning between 14 and 55 years articulated the "Emirati-emphasized speech database (Arabic database)". Eight public "Emirati utterances that are comprehensively utilized in the "United Arab Emirates" society were uttered by every speaker. Every speaker expressed the eight sentences in each of "neutral, happy, sad, disgust, angry, and fear emotions" 9 times with a span of 2 – 5 seconds. These speakers were inexperienced to keep away from any misrepresented words. The eight sentences are tabulated in Table 1 ("the right column gives the sentences in Emirati emphasize whereas the left column shows the English version"). This dataset was gathered in two isolated and diverse sessions: "instructional training session and testing session".

The captured dataset was recorded in the "College of Communication, University of Sharjah, United Arab Emirates". The database was caught by a "speech acquisition board utilizing a 16-bit linear coding analog-to-digital converter and sampled at a rate of 44.6 kHz". These signals were then "down sampled" to 12 kHz. Samples of speech signals were "pre-emphasized" and then sliced into slices of 20 ms apiece with 31.25% intersection between sequential slices.

**Table 1**
Emirati speech dataset and its English version

| No. | English Version | Emirati Accent |
|---|---|---|
| 1. | I'm leaving now, may god keep you safe. | فداعة الرحمن بترخص عنكم الحينه. |
| 2. | The one whose hand is in the water is not the same as whose hand is in the fire. | اللي ايده في الماي مب نفس اللي ايده في الضو. |
| 3. | Where do you want to go today? | وين تبون تسيرون اليوم؟ |
| 4. | The weather is nice, let's sit outdoor. | قوموا نيلس في الحوي , الجوغاوي برع. |
| 5. | What's in the pot, the spoon gets it out. | اللي في الجدر يطلعه الملاس. |
| 6. | Welcome millions, and they are not enough. | مرحبا ملايين ولا يسدن. |
| 7. | Get ready, I will pick you up tomorrow. | زهب عمرك بخطف عليك باجر. |
| 8. | Who doesn't know the value of the falcon, will grill it like a chicken. | اللي ما يعرف الصقر يشويه. |

## 4. Feature Extraction

Speech components are classified as segmental and suprasegmental based on temporal behavior [27]. Segmental components are computed for a brief timeframe traverse of 25-30 ms utilizing window strategy. However, suprasegmental components are computed over the whole utterance. Suprasegmental is a vocal effect that extends over more than one sound segment in an utterance such as pitch, stress, or juncture pattern. Suprasegmental is often used for tone, vowel length, and features like nasalization and aspiration [28]. Vector features are also categorized into two other unique groups called Low Level Descriptors (LLD) and functionals. Specifically, LLDs include prosodic and spectral features such as "fundamental frequency, energy, formants, MFCCs, Linear Prediction Cepstral Coefficient (LPCC), speaking rate, shimmer, jitter and voice quantity parameters". Functionals include their statics such as "mean, maximum, minimum, change rate, kurtosis, skewness and zero crossing rate variance" [27,28,29].

MFCC represents the transient power range of human speech. Subsequently, to get the exceptional elements of human voice, MFCC is better than other essential element vectors [27]. MFCC depends on the straight cosine change of the log control range of the direct Mel size of recurrence. Since recurrence groups are similarly dispersed in Mel recurrence, they can represent the human voice more exact. Mel frequency (m) is calculated from normal frequency (*f*) by using the formula [30],

$$m = 2595 \log (1 + f / 700) \qquad (1)$$

Mel frequency wrapping is useful for a perfect illustration of speech and effective speaker identification. Mel frequency implementation can be summarized in the following four steps:

1. The signal is being framed into 25 ms standard pattern, then the frame length for a 16 kHz input signal S(n) = 0.025 × 16000 = 400 samples. By the method of framing, S(n) is converted into $S_i(n)$, where *i* represents the number of frames. Taking the Fourier transform of the framed signal $S_i(n)$, $S_i(k)$ can be obtained as [31],

$$S_i(k) = \Sigma_{n=1}^{k} S_i(n)\, h(n) e^{-j2\pi kn/N} \qquad 1 \leq k \leq N \qquad (2)$$

where *h(n)* is the impulse response of the Hamming window and *k* is the DFT length.

The power spectral estimate of signal $S_i(n)$ is,

$$P_i(k) = \frac{1}{N} [S_i(k)]^2 \qquad (3)$$

2. Enumerate Mel spaced filter bank. This is a set of 25 triangular filters that can be applied to the periodogram power spectral estimate which indicates the energy level in each filter bank.

3. Calculate the Log values of the 26 filter bank energies of step 2.

4. Calculate the Discrete Cosine Transform (DCT) of the 26 filter bank energies of step 2 to get Spectral Coefficients.

## 5. Classification

Speaker identification uses a diversity of pattern recognition procedures to form several sophisticated classifiers such as "Gaussian Mixture Model (GMM), Hidden Markov Model (HMM), Support Vector Machine (SVM), Multilayer Perceptron (MLP) and Decision Tree (DT)". Among all classifiers, many researchers used GMM as a classifier for speaker identification and language recognition since GMM offers text-independent, robust, computationally efficient classification. Also, GMM gives better approximation for arbitrarily shaped densities [32].

### 5.1. GMM Model Description

Fig.1 shows Gaussian mixture density model as the weighted sum of *M* component densities. The Gaussian Mixture Density is defined as [32],

$$P(\bar{x}|\lambda) = \sum_{i=1}^{M} P_i b_i(\bar{x}) \qquad (4)$$

where $\bar{x}$ is the D-dimensional random vector, $b_i(\bar{x})$ represents the component densities for i =1, …, M. The component density can be defined as,

$$b_i(\bar{x}) = \frac{1}{2\pi^{D/2}|\Sigma_i|^{1/2}} \exp\left\{\frac{-1}{2}(\bar{x}-\bar{\mu})'\Sigma_i^{-1}(\bar{x}-\bar{\mu}_i)\right\} \tag{5}$$

The Gaussian mixture density parameters: mean $\bar{\mu}_i$, covariance $\Sigma_i$ and the mixture weights $P_i$ can be collectively represented and is referred as the GMM tag,

$$\lambda = \{P_i, \bar{\mu}_i, \Sigma_i\} \qquad \text{where i} = 1\ldots M \tag{6}$$

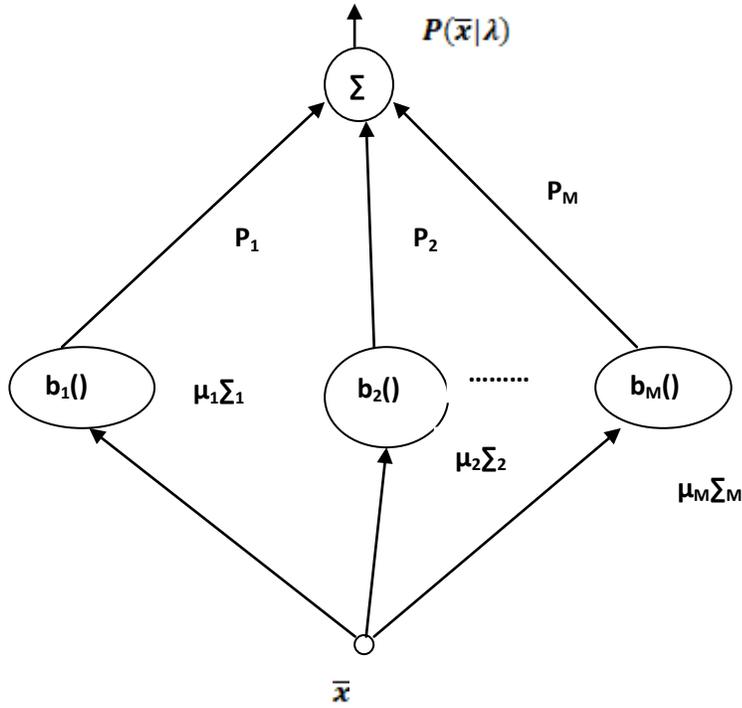

**Fig. 1.** GMM model

## 5.2. GMM Based Estimation

For speaker identification, the feature vectors are extracted from the test speech signals. Then, the feature vector sequences are divided into overlapping segments of $T$ feature vectors. The training procedure is detailed in the following steps:

1. GMM training is initialized with the initial model $\lambda$.

2. Compute the new model $\bar{\lambda}$, there by, $p(X|\bar{\lambda}) \geq p(X|\lambda)$.

3. Repeat the process to obtain the convergence,

$$p(i|\vec{x}_t, \lambda) = \frac{p_i b_i(\vec{x}_t)}{\sum_{k=1}^{M} p_k b_k(\vec{x}_t)} \tag{7}$$

Mixture weights are defined as,

$$\bar{p}_i = \frac{1}{T}\sum_{t=1}^{T} p(i|\vec{x}_t, \lambda) \tag{8}$$

Means are given by,

$$\vec{\bar{\mu}}_i = \frac{\sum_{t=1}^{T} p(i|\vec{x}_t, \lambda)\vec{x}_t}{\sum_{t=1}^{T} p(i|\vec{x}_t, \lambda)} \tag{9}$$

Variance is defined as,

$$\bar{\sigma}_i^2 = \frac{\sum_{t=1}^{T} p(i|\vec{x}_t, \lambda)x_t^2}{\sum_{t=1}^{T} p(i|\vec{x}_t, \lambda)} - \bar{\mu}_i^2 \tag{10}$$

Speaker set $S = \{1, 2,\ldots, s\}$ is represented by GMM's $\lambda_1, \lambda_2, \ldots, \lambda_s$. Then, the speaker model is defined as,

$$\hat{S} = \arg\max P(\lambda_K|X) = \arg\max \left(\frac{P(X|\lambda_K)P(\lambda_k)}{p(X)}\right) \tag{11}$$

### 5.3. Deep Neural Network

One of the most powerful machine learning tools in the last few decades is Neural Network. So, Deep Neural Network (DNN) has also a perfect and successful entry into the field of audio signal processing. DNN was introduced into the field of speaker identification as a successor of Automatic Speech Recognition (ASR) which was a comprehensive success [19].

DNN, as an application in speaker identification, has two major methods:

1. DNN for speech feature extraction frame by frame.
2. DNN as a classifier.

In the first method, the features of DNN can be derived immediately from the DNN output posterior probabilities merged with conventional features Perceptual Linear Predictive (PLP) or MFCC. DNN is trained for a specific task, where the features are taken from a narrow hidden layer compacting the related data into low dimensional feature vectors. In another approach, it is found that the great dimensional output of one of the hidden layers can be converted to features using a dimensionality decrease procedure such as Principal Component Analysis (PCA) [30]. DNN trained to frame by frame classification for feature extraction in speaker identification has only limited success. In some studies, DNN has been spotted as an effective classifier [19], [33].

In this work, the DNN that has been used is called a Convolutional Neural Network (CNN) with multiple layers of hidden units between its inputs and outputs. The CNN model is trained using an autoencoder. The hidden layer inputs $x_j$ with a Rectified Linear Unit ReLU activation function and converts it to a scalar state $y_j$ which is then forwarded to the next stage as shown below [34],

$$f(x) = \begin{cases} 0 & x < 0 \\ x & x \geq 0 \end{cases} \qquad (12)$$

In this research, DNN used four hidden layers with 128 rectified linear hidden units and gradient descend method has been used to learn the weights in DNN. The trained DNN produces probability distribution *P* over all emotions. Then, the decision block selects the particular model having the highest probability value.

The acoustic features may carry more information such as speaker details, emotional details, interference from surface reflections, reverberations and so on. Hence, the results based on a single classifier system are not sufficient to bridge the gap between the human-computer interface and intelligent human-computer interface. The proposed cascaded GMM-DNN classifier aims to improve the classification step with reduced computational complexity.

For speaker identification, the feature vectors are extracted from the test speech signals. Then, the feature vector sequences are divided into overlapping segments of *T* feature vectors. GMM tags for each speaker in each emotional talking condition are being created. For each speaker, they provide a unique tag to each emotion, such as neutral, sad, happy, disgust, angry, and fear. During the test phase, the log likelihood distance between the voice query and each of the GMM tag is compared and a new vector of features is created. This new vector of features is the input of the DNN classifier. The Rectified Linear Unit DNN (ReLU DNN) with four hidden layers have been used, each with 128 rectified linear units which provide the final decision.

The classification step using DNN is shown in Fig. 2. The training target of DNN is the correct speaker identification. DNN is used to perform the classification task followed by GMM. In this work, the CNN architecture shown in Fig. 2, consists of 4 convolution layers. Each convolution layer is followed by a max-pooling layer. Fully connected layers combine the output of max-pooling layers and GMM tags to achieve the classification. This work uses the concept of the Ideal Binary Mask at the output. The sum of probabilities is equal to one when the results of GMM and DNN are identical while the other speakers have zero probabilities.

## 6. Speaker Identification Algorithm based on Cascaded GMM-DNN and the Experiments

The block diagram of the proposed classifier is shown in Fig. 3. Emirati Speech Database (ESD) includes eight unique sentences prompted at six different emotions by twenty five female and twenty five male native Arabic speakers. During the training stage, the first four sentences of ESD are used while in the evaluation stage, the last four utterances of ESD are utilized. During the training phase, the MFCC features are extracted for all the training data and these features are fed to the GMM based preliminary classifier. The GMM based classifier generates the GMM tag for each speaker in different emotions.

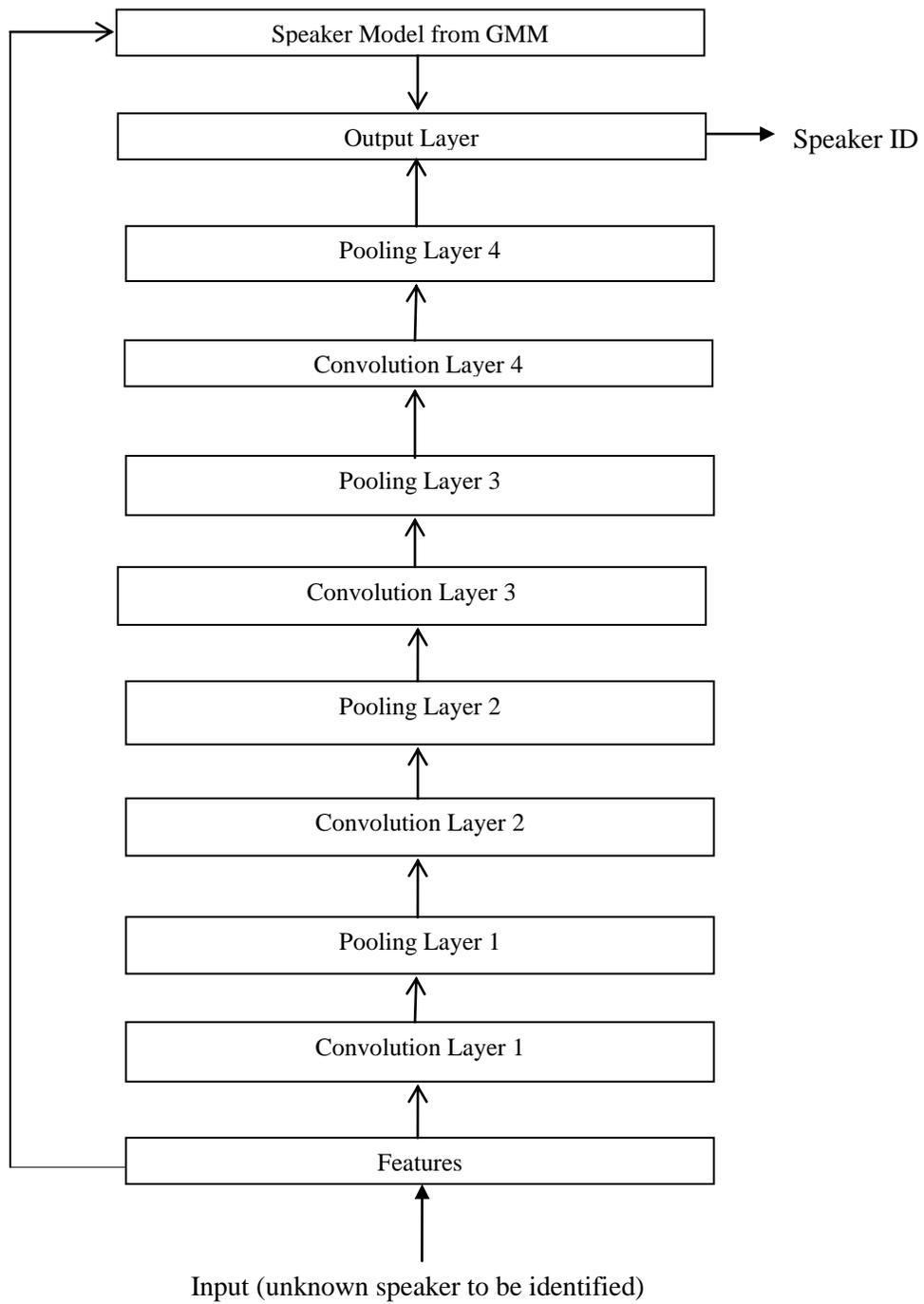

**Fig. 2.** Topology of DNN classification

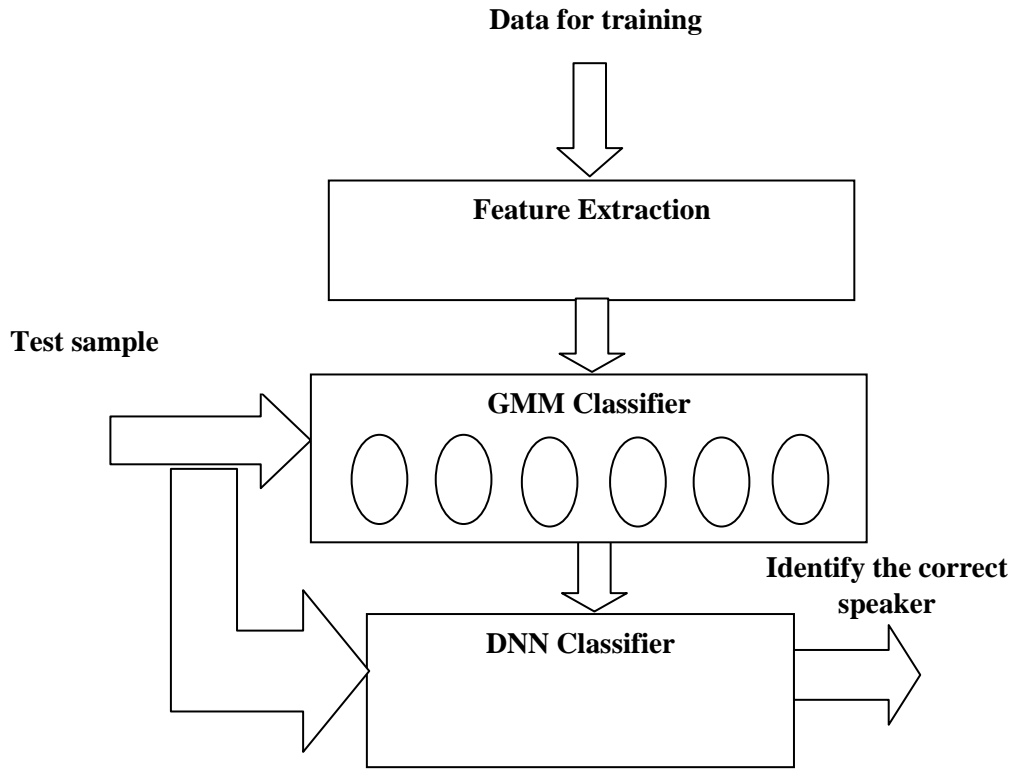

**Fig. 3.** Block diagram of the proposed classifier

During the "evaluation phase", the "log likelihood distance" between the voice query and each of the GMM tags is compared for each of the emotional state and, thereby, generates a new vector of features which is the input of the DNN classifier that produces the final decision. In the training phase, the total number of speech samples used is10,800 (50 speakers ✕ the earliest 4 utterances ✕ 9 repetitions ✕ 6 emotions). In the test phase, the total number of speech samples used is10,800 (50 speakers ✕ the latest 4 utterances ✕ 9 repetitions ✕ 6 emotions). Fig. 4 shows the detailed view of the cascaded GMM-DNN classifier stage.

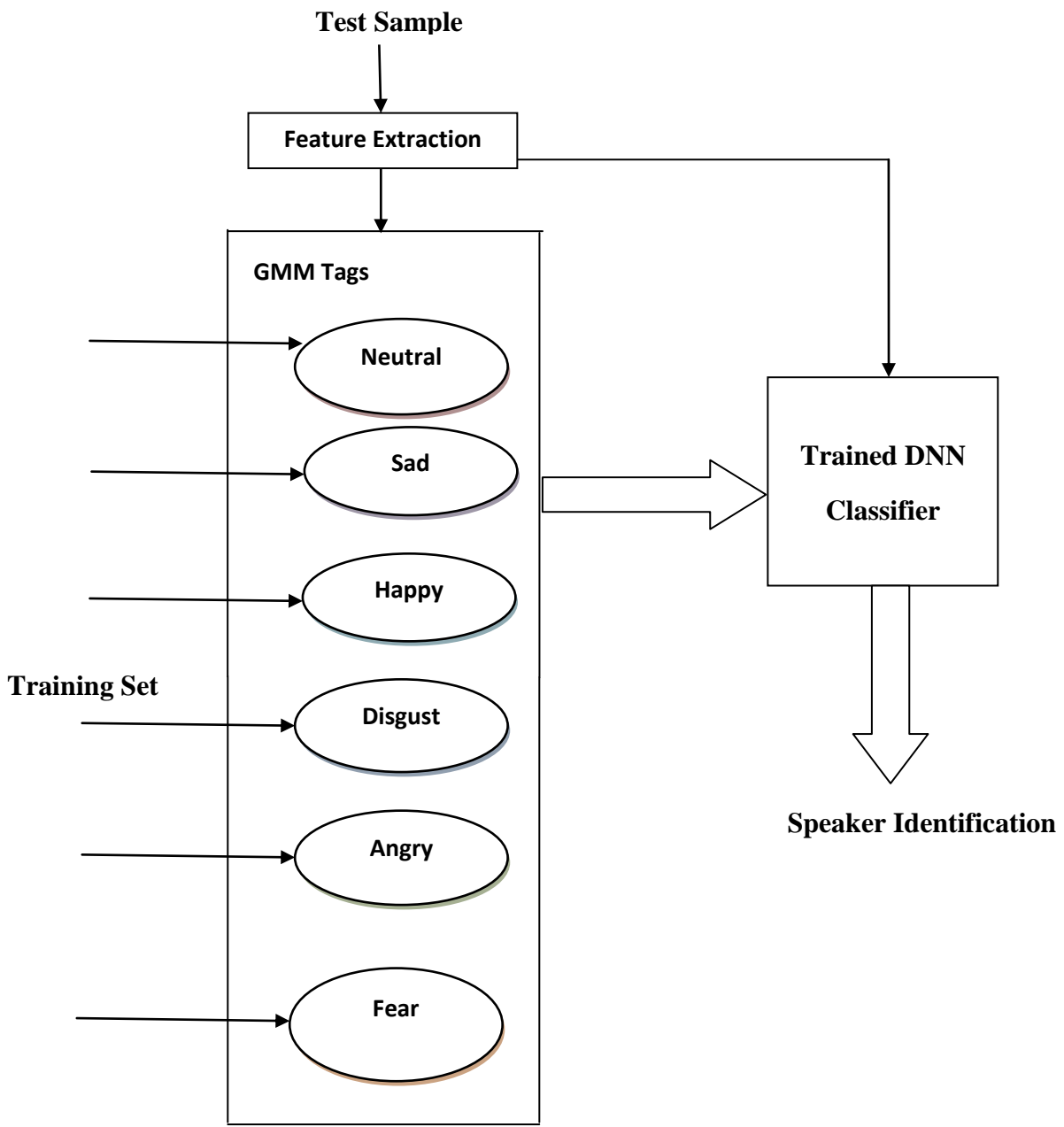

**Fig. 4.** GMM-DNN Classification

## Results and Discussion

The current research suggests, executes and evaluates a novel classifier in light of cascaded GMM-DNN for speaker identification in each of neutral and emotional environments. To survey the proposed classifier, speaker identification performance has been independently compared in light of each of SVM and MLP-based speaker identification in neutral and emotional environments. Average "text-independent speaker identification in each of neutral and emotional environments" utilizing the gathered dataset in view of each of cascaded GMM-DNN, SVM and MLP is shown in Fig. 5. Speaker identification performance has been calculated as:

$$\text{SID Performance} = \frac{\text{Total number of times the unknown speaker has been identified correctly}}{\text{Total number of trials}} \times 100\% \quad (13)$$

Fig. 5 shows that every model functions practically perfect in the neutral environment. In such environment, the average speaker identification performance based on GMM-DNN demonstrates a growth of 2.7% and 3.3% over that based on SVM and MLP, respectively. In addition, this figure demonstrates that the execution in emotional talking environments indicates a significant performance improvement based on GMM-DNN over that based on each of SVM and MLP. Despite that, the framework performance is poor while examining the execution in "Angry" emotion.

"A statistical significance test" has been carried out to indicate whether speaker identification rate variations (speaker identification rate using GMM-DNN and that using each of SVM and MLP in each of neutral and emotional environments) are actual or just due to statistical variabilities. This test has been performed using the "Student's *t* Distribution test" as given by the accompanying equation,

$$t_{1,2} = \frac{\bar{x}_1 - \bar{x}_2}{SD_{pooled}} \qquad (14)$$

where "$\bar{x}_1$ is the mean of the first sample of size *n*, $\bar{x}_2$ is the mean of the second example of a similar size, and $SD_{pooled}$ is the pooled standard deviation of the two samples given as",

$$SD_{pooled} = \sqrt{\frac{SD_1^2 + SD_2^2}{2}} \qquad (15)$$

where "$SD_1$ is the standard deviation of the first sample of size *n* and $SD_2$ is the standard deviation of the second sample of equal size" [35].

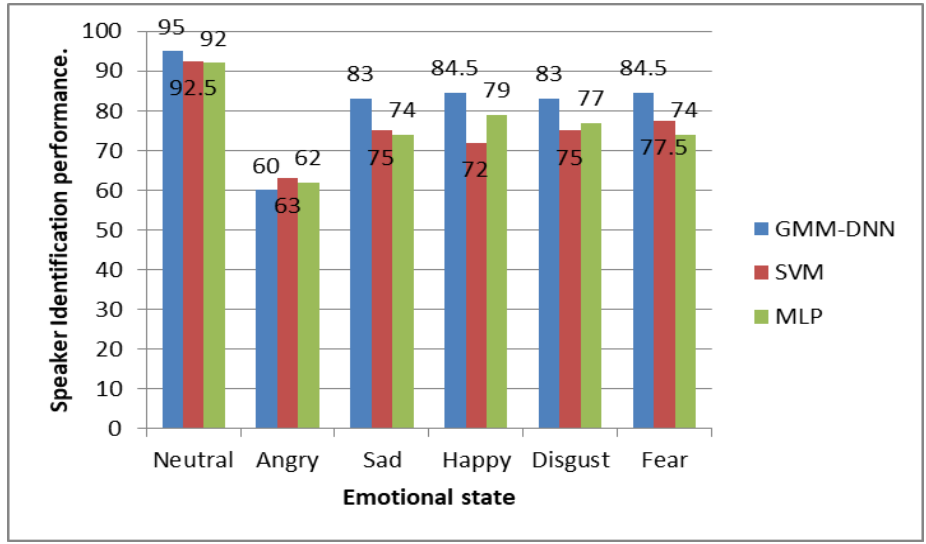

**Fig. 5.** Average speaker identification performance evaluation using ESD based on each of GMM-DNN, SVM and MLP

In this work, the computed *t* values between cascaded GMM-DNN and each of SVM and MLP in neutral and emotional talking conditions using the ESD are given in Table 2. In view of this table, each figured *t* value in neutral talking condition is less than the "tabulated critical value $t_{0.05}$= 1.645 at 0.05 significant level". In this talking condition, the predominance of GMM-DNNs over each of the other two classifiers is insignificant since the other acoustic models perform well in such a talking condition as shown in Fig. 5. In the emotional talking conditions (except for "Angry" emotion), each calculated *t* value is higher than the "tabulated critical value $t_{0.05} = 1.645$". Hence, speaker identification rate using GMM-DNN leads that using each of SVM and MLP in emotional talking conditions with the exception of angry emotion. The analysis shows that our proposed classifier gives significant improvement in emotional talking conditions (except for "Angry" emotion) because the calculated *t* values are greater than the "tabulated critical value $t_{0.05} = 1.645$". On the other hand, our proposed classifier does not show significant enhancement in neutral condition since the computed *t* values are smaller than the tabulated critical value.

**Table 2**

Calculated *t* values between GMM-DNN and each of SVM and MLP in neutral and emotional talking environments utilizing the ESD

| $t_{1,2}$ | Calculated *t* value | | | | | |
|---|---|---|---|---|---|---|
| | Neutral | Sad | Happy | Disgust | Fear | Angry |
| $t$ GMM-DNN, SVM | 1.157 | 1.832 | 1.791 | 1.835 | 1.730 | 1.125 |
| $t$ GMM-DNN, MLP | 1.135 | 1.964 | 1.823 | 1.690 | 1.680 | 1.225 |

Four additional experiments have been autonomously accomplished to assess speaker identification rate achieved in neutral and emotional conditions using the cascaded GMM-DNN. The four experiments are:

1) **Experiment 1:** The SUSAS English speech dataset has been used to evaluate the three classifiers. The main objective of SUSAS dataset was firstly to study speech recognition in neutral and stressful environments [36]. Angry talking condition and shouted talking conditions are utilized as alternatives since in normal life it is difficult to separate them [37]. Thirty different utterances of seven speakers in each of neutral and angry talking environments have been chosen in this experiment. "Average speaker identification performance" using SUSAS dataset based on GMM-DNN, SVM and MLP in neutral and angry environments is shown in Fig. 6. This figure implies that the proposed framework performance is low while examining the execution of the emotion "Angry". This figure shows speaker identification performance degradation of 12.0% and 23.2% from our proposed framework in light of MLP and SVM, respectively.

Table 3 shows analysis of speaker identification performance in angry condition of the SUSAS database based on six different classifiers. From the system implementation outcomes, it is apparent that the proposed classifier yields improved "speaker identification performance" in angry condition competed to the four classifiers and models (SVM, Genetic Algorithm, Vector Quantization and MLP). On the other hand, the novel GMM-DNN classifier shows degraded "speaker identification performance" in the same talking condition competed to "Third-Order Circular Suprasegmental Hidden Markov Model (CSPHMM3)". This analysis demonstrates the need for a vital classifier which improves the performance in angry condition.

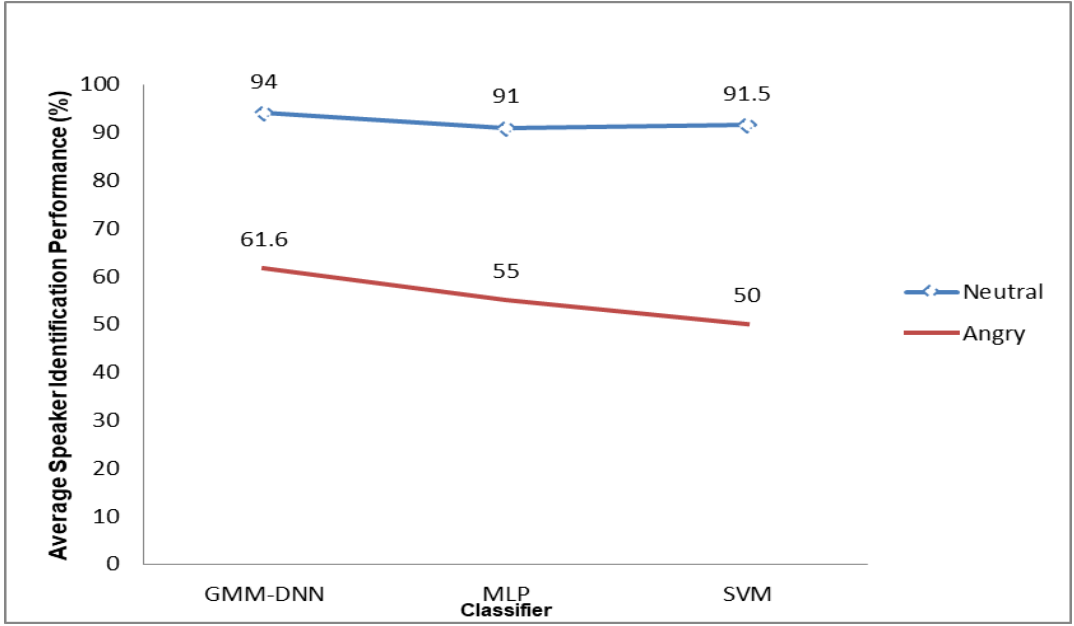

**Fig. 6.** Average speaker identification performance utilizing SUSAS database in neutral and angry conditions

**Table 3**
Speaker identification performance in angry talking condition using SUSAS dataset based on different classifiers

| Classifier | CSPHMM3 | SVM | GMM-DNN | GA | VQ | MLP |
|---|---|---|---|---|---|---|
| Speaker Identification Performance (%) | 81.8 | 50.0 | 61.6 | 58.3 | 56.6 | 55.0 |

2) **Experiment 2**: A casual, subjective assessment of GMM-DNN using the ESD has been completed with ten nonprofessional Arabic audience members (human judges). Altogether 120 speech samples (5 speakers × 4 sentences × 6 emotions) are utilized in this experiment. Assessment phase proceeds with each listener independently advised to recognize the unknown speaker uttering in neutral and emotional environments (totally two different and independent environments). The graphical representation of this performance analysis based on the subjective evaluation is demonstrated in Fig.7. The graph shows that human listener performance is close to the proposed classifier performance except for angry emotion.

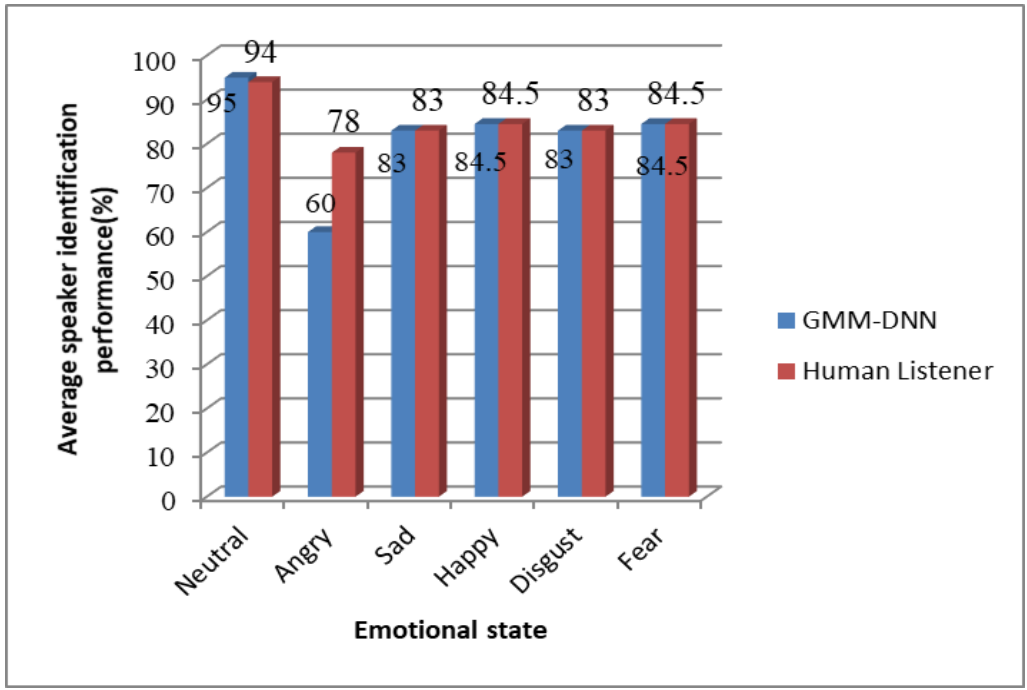

**Fig. 7.** Speaker identification performance analysis based on GMM-DNN and human listeners

3) **Experiment 3:** During the evaluation phase, test data are mixed with some interference signals in a ratio 2:1 and the results obtained are shown in Table 4. Randomly selected 3 speech samples from each emotion mixed with interference signal are used as the test data, thereby, the final outputs are compared with other classifiers. Our outcomes illustrate that speaker identification performance does not decrease in the presence of noise. This is achieved by the use of the GMM speaker identification tag. This creates a mask for the speaker data from other feature vectors. Average speaker identification performance using the distorted data shows 4% - 5% degradation compared to results obtained using the normal data in neutral and emotional talking conditions. Table 4 also shows that the results attained based on GMM-DNN are greater than those achieved based on each of SVM and MLP classifiers. From the proposed classifier implementation outcomes, it is apparent that GMM-DNN gives enhanced results for angry talking condition using the distorted data. These results show that the proposed classifier yields better results (76.8%) than each of SVM (63.0%) and MLP (62.7%) classifiers in noisy talking environments.

**Table 4**
Performance analysis of speaker identification using distorted data/normal data

| Emotion | Speaker Identification Performance (%) | | | | | |
|---|---|---|---|---|---|---|
| | GMM-DNN | | SVM | | MLP | |
| | Distorted data | Normal data | Distorted data | Normal data | Distorted data | Normal data |
| Neutral | 90 | 95 | 88.2 | 92.5 | 85 | 92 |
| Angry | 55 | 60 | 50 | 63 | 55 | 62 |
| Sad | 79 | 83 | 60 | 75 | 62 | 74 |
| Happy | 80 | 84.5 | 58 | 72 | 60 | 79 |
| Disgust | 78 | 83 | 59.5 | 75 | 55 | 77 |
| Fear | 79 | 84.5 | 62.3 | 77.5 | 59 | 74 |
| **Average** | **76.8** | **81.7** | **63.0** | **75.8** | **62.7** | **76.3** |

A casual subjective assessment of GMM-DNN performance in distorted speech has been completed with ten non professional Arabic audience members. Altogether 120 distorted speech samples (5 speakers × 4 sentences × 6 emotions) are used in this experiment. In the test phase, each speaker is advised to recognize the unknown speaker talking in neutral and emotional talking environments. The graphical representation of the performance analysis is shown in Fig. 8. This figure illustrates that, for distorted data, speaker identification performances based on GMM-DNN and human listeners are very close.

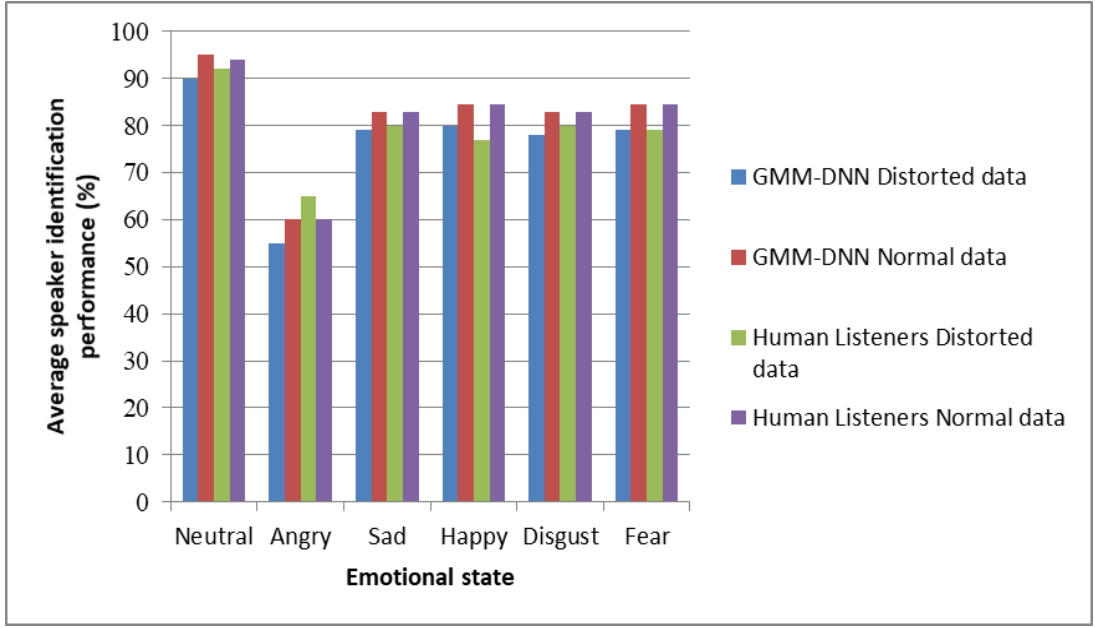

**Fig. 8.** Speaker identification performance analysis based on GMM-DNN and human listeners in normal and distorted data

4) **Experiment 4:** This experiment has been conducted to show the relevance of the proposed GMM-DNN as a classifier and to compare it to each of GMM alone and DNN alone. Average "speaker identification performance" using the ESD dataset based on each of GMM-DNN, GMM alone, and DNN alone is shown in Table 5. This table yields 17.9% and 7.2% improvement rate in speaker identification performance based on the proposed classifier over that based on GMM alone and DNN alone, respectively. These results evidently demonstrate that the cascaded GMM-DNN is superior to each of GMM alone and DNN alone. The best approximation property of GMM classifier and the fine tuning used by the DNN classifier is aptly combined in this work to achieve improvements in "average speaker identification performance". The proposed classifier offers a robust and computationally efficient novel classification technique for "speaker identification in emotional environments".

**Table 5**

Performance analysis of speaker identification based on each of GMM-DNN, GMM alone, and DNN alone using the ESD

| Classifier | "Neutral" | "Angry" | "Sad" | "Happy" | "Disgust" | "Fear" | Average |
|---|---|---|---|---|---|---|---|
| GMM-DNN | 95.0 | 60.0 | 83.0 | 84.5 | 83.0 | 84.5 | 81.7 |
| GMM | 90.0 | 53.0 | 61.3 | 68.2 | 70.9 | 72.3 | 69.3 |
| DNN | 92.0 | 58.0 | 75.1 | 79.3 | 77.2 | 75.4 | 76.2 |

5) **Experiment 5:** This experiment has been conducted to show the relevance of the proposed GMM-DNN as a classifier to enhance speaker identification performance in emotional environments and to compare it with other classifiers in the literature [19], [21], [22]. Matejka *et.al* [19] studied utilizing Deep Neural Network Bottleneck (DNN-BN) features together with MFCCs in the task of i-vector-based speaker recognition. Richardson *et.al* [21] presented the application of single DNN for both speaker recognition and language recognition using the "2013 Domain Adaptation Challenge speaker recognition (DAC13)" and the "NIST 2011 Language Recognition Evaluation (LRE11)" benchmarks. Ali *et.al* [22] studied the use of features from distinct levels of Deep Belief Network (DBN) to quantize the audio data into vectors of "audio-word counts". Table 6 demonstrates speaker identification performance based on each of GMM-DNN and the three different prior work using the ESD. This table yields 5.8%, 3.9%, and 5.6% improvement rate in speaker identification performance based on the proposed classifier over that based on DNN-BN [19], single DNN [21], and DBN [22], respectively. Hence, GMM-DNN offers a robust and computationally efficient novel classification technique for "speaker identification in emotional environments".

**Table 6**

Performance analysis of speaker identification based on each of GMM-DNN and three different prior work using the ESD

| Method | "Neutral" | "Angry" | "Sad" | "Happy" | "Disgust" | "Fear" | Average |
|---|---|---|---|---|---|---|---|
| GMM-DNN | 95.0 | 60.0 | 83.0 | 84.5 | 83.0 | 84.5 | 81. 7 |
| DNN-BN | 94.5 | 60.1 | 75.1 | 79.5 | 77.4 | 76.5 | 77.2 |
| Single DNN | 93.5 | 59.5 | 79.1 | 80.5 | 79.5 | 79.6 | 78.6 |
| DBN | 92.9 | 64.5 | 77.0 | 75.0 | 76.0 | 79.0 | 77.4 |

## 8. Concluding Remarks

Novel cascaded GMM-DNN based-classifier has been designed, implemented and evaluated to enhance "text-independent speaker identification performance in emotional talking environments". This work shows that the proposed classifier yields better results than those mostly used each of SVM and MLP based classifiers. In addition, the cascaded GMM-DNN draws higher speaker identification performance than each of GMM alone and DNN alone. Furthermore, GMM-DNN leads each of DNN-BN, single DNN, and DBN for speaker identification in emotional environments. The proposed system also gives better performance in distorted speech signals. The algorithm based on GMM tag based-feature vector reduction helps minimizing the complication of the DNN classifier, thereby, improving system performance. The proposed classifier gives better results even in the presence of interference.

Our proposed classifier fails to improve speaker identification rate when speakers speak angrily. A further extensive study and research is in progress to upgrade speaker identification rate analysis in such condition. Our future work aims to enhance the system performance in angry and noisy talking environments with reduced computational complexity.

**Compliance with Ethical Standards**

Ismail Shahin, Ali Bou Nassif, and Shibani Hamsa would like to thank "University of Sharjah" for funding their work through the two competitive research projects entitled "Emotion Recognition in each of Stressful and Emotional Talking Environments Using Artificial Models", No. 1602040348-P, and "Capturing, Studying, and Analyzing Arabic Emirati-Accented Speech Database in Stressful and Emotional Talking Environments for Different Applications", No. 1602040349-P.

**Conflict of Interest**: The authors declare that they have no conflict of interest.

**Informed consent**: This study does not involve any experiments on animals.

**Statement of Ethics**: The authors have permission from the University of Sharjah to collect speech database from UAE citizens based on the two competitive research grants entitled "Emotion Recognition in each of Stressful and Emotional Talking Environments Using Artificial Models", No. 1602040348-P, and "Capturing, Studying, and Analyzing Arabic Emirati-Accented Speech Database in Stressful and Emotional Talking Environments for Different Applications", No. 1602040349-P.

**Consent of Parents' of Minors**: This study includes very few speakers who are less than 18 years old. A consent from the minors' parents was provided before conducting the experiments.